\documentstyle[floats,aps,psfig]{revtex}
\begin{document}

\flushbottom

\twocolumn[\hsize\textwidth\columnwidth\hsize\csname @twocolumnfalse\endcsname

\title{Quantum atomic delocalization vs. structural disorder 
       in amorphous silicon}
\author{Carlos P. Herrero }
\address{Instituto de Ciencia de Materiales, C.S.I.C.,
         Campus de Cantoblanco, 28049 Madrid, Spain \\ and \\
	 Institut f\"ur Theoretische Festk\"orperphysik,
	   Universit{\"a}t Karlsruhe, D-76128 Karlsruhe, Germany}
\date{\today}
\maketitle

\begin{abstract}
 Quantum effects on the atom delocalization in amorphous silicon have
been studied by path-integral
Monte Carlo simulations from 30 to 800 K.
The quantum delocalization is appreciable vs. topological disorder, as seen
from structural observables such as the radial distribution function
(RDF). At low temperatures, the width of the first peak in the RDF
increases by a factor of 1.5 due to quantum effects.
The overall anharmonicity of the solid vibrations at finite 
temperatures in amorphous silicon is clearly larger
than in the crystalline material. Low-energy vibrational modes
are mainly located on coordination defects in the amorphous 
material.  \\
\end{abstract}
\vskip2pc]

\narrowtext

 Structural and dynamical aspects of amorphous solids
have been studied for many years by several experimental
and theoretical techniques \cite{el90,cu87}. 
In spite of the intense work devoted to determine the structure 
of amorphous materials at an atomic level, there are still many
open questions concerning the short and intermediate length
scales in these materials \cite{mo97}.
In this context, amorphous silicon,
apart from its technological interest, 
is a model system to analyze energy-minimized
structures in glassy solids \cite{ba96}, as well as 
low-energy excitations in disordered semiconductors \cite{li97}.

In the last decade, computer simulations of semiconductors with
topological disorder have been performed, in order to obtain
insight into different structural and dynamical properties
of these materials. 
Early simulations \cite{wo85} of amorphous silicon (a-Si) 
were based on the continuous random network
model (see ref. \onlinecite{el90}), and  later, 
 most simulations of this material have been
carried out by molecular dynamics (MD) with empirical interatomic
potentials \cite{kl87,kl88,bi88,co93}.
$Ab \; initio$ MD simulations with the Car-Parrinello method
were also carried out for a-Si, and their
results for the atomic structure, the phonon spectrum, and the
electronic properties were in good agreement with experimental
data \cite{ca88}.
These simulations, however, treat the silicon nuclei as
classical particles, thus neglecting typical quantum effects
such as tunneling or zero-point motion. The
quantum delocalization of the atomic nuclei in solids is
usually taken into account through a harmonic approximation, 
but such an approach is in principle not valid for amorphous solids,
due to the presence of highly anharmonic low-energy vibrations in these
materials \cite{bu88,gi93}.
In last years, $ab \; initio$ path-integral MD simulations have
become possible for small molecules, though they are still
hardly feasible for detailed studies of solids \cite{ma95}.

An important point, when considering average structural properties
of amorphous materials, is the question whether quantum effects
can be important vs. structural disorder. In particular, one may
ask whether the radial distribution function (RDF) obtained in
classical
simulations will be changed by introducing quantum effects, with
the consequent quantum delocalization of the atoms, which
can be noticeable mainly  at low temperatures.
Thus, two factors appear to compete in the broadenig of the RDF
peaks at low $T$: structural disorder and quantum delocalization.
As a first approach, one expects that for amorphous materials 
with heavy atoms (small zero-point delocalization), the
structural disorder will broaden the peaks more than the
zero-point motion, and the contrary will happen for disordered
materials with light atoms.

  The purpose of this paper is twofold: First, to analyze whether
the average structural properties of amorphous silicon are 
affected by quantum effects, 
and second, to check the validity of a harmonic approximation to
obtain average quantities (energy, mean atom delocalization)
at finite temperatures.
With these goals, amorphous silicon has been  studied
 by path-integral (PI)
Monte Carlo (MC) simulations in a temperature range from
30 to 800 K.
 This computational technique is now well established
as a tool to study many-body problems in which anharmonic
effects can be non-negligible. It
is based on the Feynman path-integral formulation
of statistical mechanics, according to which the partition function
of a quantum system can be evaluated through  a discretization
of the density matrix along cyclic paths, composed of a
finite number of imaginary-time steps \cite{fe72}.
This Trotter number, $N$, causes the appearance in the simulations
of $N$ `replicas' for each quantum particle. Such replicas are
treated as classical particles, and the corresponding partition
function is sampled by the Metropolis method \cite{me53}.
More details on this simulation procedure can be found 
elsewhere  \cite{gi88,ra93}.

\begin{figure}
\centerline{\psfig{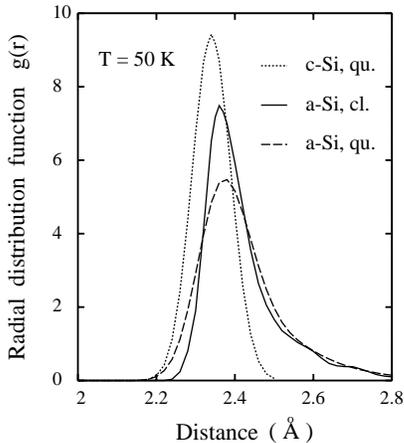}}
\vspace*{-2.5cm}
\caption{
Radial distribution function in the region from 2 to 2.8 \AA,
as obtained from Monte Carlo simulations of a-Si at 50 K:
classical (continuous line) and quantum (dashed line).
Results of quantum simulations for c-Si are plotted as a dotted line.
}
\end{figure}

The interatomic interaction has been  modelled by the Stillinger-Weber
(SW) potential, \cite{st85} which was employed
 in earlier PI MC simulations
 of crystalline silicon (c-Si), giving results
(total energy, quantum delocalization of the Si atoms) 
in good agreement with those derived from experiment \cite{ra93}.
This interatomic potential was also shown to be adequate to
simulate structural and dynamical properties of a-Si in classical
MD simulations \cite{kl87,kl88}.
A cubic cell for a-Si with a size length of 16.6 \AA,
 and including 216 atoms,
 was generated by using a simulated annealing process
similar to that described in refs. \onlinecite{kl87}
and \onlinecite{kl88}, where an
amorphous structure was generated by MD simulations, via slowly   
cooling from the melt.  In contrast
with these works, we have carried out the simulated annealing
by using classical MC simulations with the same SW potential 
employed later for the quantum simulations. In this process, the
temperature $T$ was reduced from 2400 to 200 K in 15 steps, involving
a total of 3 $\times 10^6$ MC steps (MCS).
Several simulation cells were generated by the same procedure
to check that the average results obtained did not depend on the particular
cell under consideration. The structural characteristics of our
generated cells are similar to those found in earlier MD simulations
with the SW potential,
in particular the radial distribution function (RDF) and the fraction
of five-fold coordinated atoms ($\sim$ 20\%).
Once the simulation cell for a-Si was generated, we carried out
PI MC simulations in a temperature range from 30
to 800 K.  The Trotter number, $N$, was made temperature-dependent
($N \sim 2000/T$), in order to keep the numerical error caused by the
discretization of the path integrals lower than $\pm$ 1meV per 
Si atom \cite{ra93}. 
At each temperature, 5000 MCS were carried out for system  
equilibration, and 15,000 to calculate the average variables of our
simulation cell. Each MCS included an update of the coordinates of all
replicas for each Si nucleus, as well as an attempt to move the
center-of-gravity of the cyclic path associated to each nucleus.
The results of our simulations are compared to those found
for crystalline silicon with the same simulation method
and the same number of atoms in the simulation cell.
In order to find out the importance of quantum effects,
classical MC simulations were also carried out for amorphous and
crystalline silicon in the same temperature range.

\begin{figure}
\centerline{\psfig{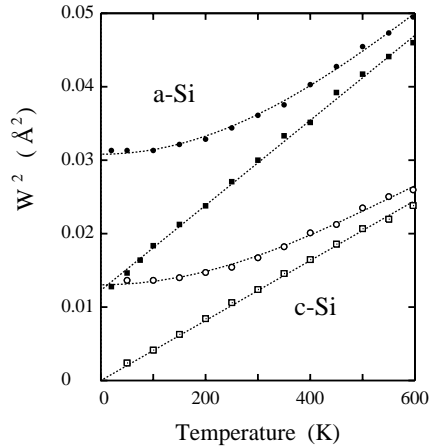}}
\vspace*{-2.5cm}
\caption{ Squared FWHM of the first peak in the radial distribution
function, as a function of temperature,
 as derived from Monte Carlo simulations: open squares,
classical c-Si; open circles, quantum c-Si; black squares,
classical a-Si; black circles, quantum a-Si. Lines are guides
to the eye.
}
\end{figure}

Traditionally, the RDF has been a useful means to characterize
the structure  of amorphous materials.
In fig.\,1 we present the RDF for a-Si obtained from our classical 
(continuous line) and quantum (dashed line)  Monte Carlo simulations
at 50 K, in the range from 2 to 2.8 \AA. 
 For comparison, we give also the RDF found for c-Si in quantum
simulations at the same temperature (dotted line).
In the classical simulations of amorphous silicon,
the first peak in the RDF has its maximum at a distance
$r_m =$ 2.36 \AA, close to the nearest-neighbor distance
in c-Si. However, in these simulations the average Si--Si distance
in a-Si is found to be somewhat higher (2.43 \AA), as a
consequence of the peak asymmetry (we employ a
nearest-neighbor cutoff distance of 2.9 \AA). 
For quantum a-Si, we find the peak-maximum slightly
shifted to higher distances ($r_m$ = 2.37 \AA), but the
mean Si--Si distance is the same as for `classical'
a-Si. The peak found in quantum simulations at low
temperatures is, however, clearly broader than that obtained 
in the classical case.

In fig.\,2 we show the squared full-width-at-half-maximum (FWHM,
denoted here by $W$),
derived from our MC simulations (both, classical and quantum)
for crystalline and amorphous silicon. In the classical simulations,
one finds a linear dependence of $W^2$ vs. $T$. For the crystal
(open squares),
$W$ goes to zero at low $T$, as should happen for a classical
approach. For amorphous silicon (black squares),
 $W$ converges to a finite value
at $T$ = 0 ($W_0 = 0.11$ \AA), as a consequence of 
structural disorder.
Our value of $W$ for classical simulations of a-Si at $T$ = 300 K
($W = 0.175$ \AA)
compares well with that found in earlier MD simulations
($W = 0.17 \pm 0.01$ \AA) \cite{co93}.

 The width $W$ obtained for c-Si in
quantum simulations at low temperatures (open circles) 
is due to zero-point motion of the Si atoms in the crystal.
This width is found to be similar to that
obtained for a-Si in the classical simulations (the latter being due
to structural disorder in the amorphous material).
As an important result, we find that,
at low $T$, $W^2$ for quantum a-Si (black circles) is found to be 
about 2.5 times the value corresponding to `classical' a-Si.
For distances larger than 3 \AA, 
the RDF's obtained in classical and quantum simulations coincide. 
To our knowledge, there are no experimental data on a-Si to
 compare directly
with our calculated $W$ values, but the FWHM found for amorphous
germanium  at room
temperature \cite{et82} is of the order of that obtained here.

The structural disorder present in a-Si affects the bond lenghts
and bond angles as follows:  Bond lengths are,
in average, larger in a-Si than in c-Si, 
even for Si--Si bonds between four-fold coordinated atoms.
Also, the bond-angle distribution for four-fold coordinated atoms
in a-Si is broader than in the crystalline material. In fact, for
$T = 50$ K we find a Si-Si-Si angle distribution  with a FWHM
of $22^{\circ}$ around the tetrahedral angle.
At this temperature, the corresponding
width for c-Si is $5.5^{\circ}$, as derived from our
quantum simulations.
This angle distribution is much broader for coordination defects
in a-Si, with a constant width of $\sim 70^{\circ}$, 
irrespective of temperature. 
 The longer average Si-Si distance in a-Si causes: (i) a
 decrease in the stretching frequency of the Si-Si bond, in
 particular, and a decrease in the zero-point energy $\frac{1}{2}
 \langle \hbar \omega \rangle$
 obtained from the vibrational density of states (VDOS); 
 and (ii) an increase in
 the anharmonicity of the solid vibrations, related to the
 metastability of the amorphous solid.
 The same trend is found for the effect of angle distorsions
 on the zero-point vibrational energy and
 on the anharmonicity (see below).

\begin{figure}
\centerline{\psfig{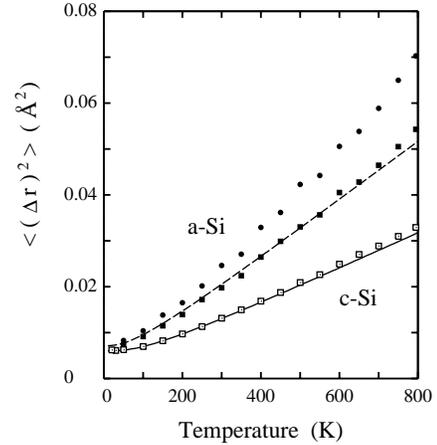}}
\vspace*{-2.5cm}
\caption{
Mean-square displacement of the atoms in crystalline and amorphous
silicon. Open squares: c-Si; black squares and circles for four-
and five-fold coordinated atoms in a-Si, respectively. The continuous
and dashed lines were calculated from the vibrational density-of-states for
crystalline and amorphous silicon, respectively.
}
\end{figure}

An important value that can be obtained directly from the quantum 
simulations is the spatial delocalization of the silicon atoms
in the material. One expects that this delocalization will be
higher for a-Si than for c-Si due to the low-energy vibrational
modes that appear in disordered materials \cite{kl88,gi93},
 and that cause an increase in the VDOS at low energies, as  
compared with the crystalline solid. In fig.\,3 we show the average
delocalization, $(\Delta r)^2$, of the Si atoms vs. $T$. Open
squares correspond to c-Si, and black squares and circles represent
the delocalization of four- and five-fold coordinated Si atoms in amorphous
silicon, respectively. The continuous and dashed lines were obtained
from the VDOS obtained for the SW potential \cite{kl88} 
by assuming a harmonic approximation.
As expected, the mean-square displacement of the four-fold 
coordinated atoms
in a-Si is clearly larger than that corresponding to c-Si.
Moreover, five-fold coordinated Si atoms (black circles in fig.\,3) 
are more delocalized 
than four-fold coordinated atoms, indicating that the former
have a larger contribution to localized low-energy vibrational modes.
This is in line with the fact that coordination defects give rise
to extra states at low frequencies \cite{bi88}.
Note that a small extra contribution of low-frequency modes to the
local VDOS for five-fold coordinated atoms, can have an 
appreciable effect on $(\Delta r)^2$ at finite temperatures,
as that shown in fig.\,3.
The results obtained for the mean-square displacement of
four-fold coordinated atoms follow closely (for both a-Si and c-Si)
the lines derived from the harmonic approximation.
Only at  $T >$ 600 K, a slight deviation of the PI MC results
toward higher values is found in both cases.

The influence of quantum effects on the atomic motion
can be quantified by calculating the correlations between atom
displacements. We define the
correlation $\rho$ for displacements of atom pairs 
by the quantity:

\begin{equation}
 \rho(r) = \frac { \langle {\bf u}_i \cdot {\bf u}_j \rangle } 
     { \sqrt { \langle  {\bf u}_i^2 \rangle  \langle {\bf u}_j^2 \rangle } } 
               \hspace{5mm}  ,
\end{equation}
where ${\bf u}_i$ (${\bf u}_j$) is the displacement of atom $i$ ($j$)
 from its equilibrium position, and $r$ is the distance between the
equilibrium positions of atoms $i$ and $j$. 
We have calculated this pair correlation as a function of $r$
in quantum and classical MC simulations.
In fig.\,4 we present the correlation $\rho(r)$ for both
amorphous and crystalline silicon at $T$ = 50 and 600 K. The dashed
and dotted lines correspond to a-Si at these temperatures.
 At $T$ = 600 K, we find that the correlation $\rho$
is $\sim$ 0.4 at a nearest-neighbor distance 
$r$ = 2.35 \AA, and decreases as the interatomic distance
increases (dotted line). 
For $r >$ 6 \AA, this correlation is nearly zero. At low $T$
(dashed line),  quantum delocalization causes a decrease in the
correlation of atom displacements.
 The black and white circles were obtained from 
quantum simulations of crystalline silicon. These points follow
 closely the corresponding lines
for amorphous silicon. This means that, in spite of the apparent
differences between the VDOS of both materials \cite{kl88},
the correlations between atom displacements do not depend 
appreciably on the degree of topological
disorder in silicon.  Note in fig.\,4 that the values obtained for
distances larger than 6 \AA\ seem to converge to a negative value.
This is a finite-size effect, since the smallest wave vector allowed
by our supercell size is $k_{min} \sim$ 0.4 \AA$^{-1}$.
In the classical simulations, one finds, for both a-Si and c-Si,
 that $\rho(r)$ is
roughly independent of temperature, and coincides with
the high-temperature limit of the quantum simulations.
In fact, in a classical harmonic approximation, $\rho(r)$
should be temperature-independent, as a consequence of the
equipartition principle.

\begin{figure}
\centerline{\psfig{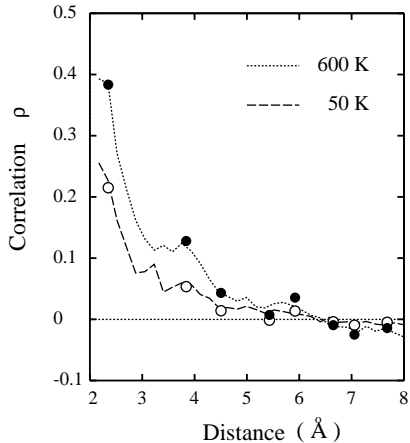}}
\vspace*{-2.5cm}
\caption{
Correlation $\rho$ between atom displacements vs. interatomic
 distance, at
two different temperatures: Open circles, c-Si at 50 K; black circles,
c-Si at 600 K; dashed line, a-Si at 50 K; dotted line, a-Si at 600 K.
Error bars are of the order of the symbol size.
}
\label{f4}
\end{figure}

    PI MC simulations can be used to estimate the overall
anharmonicity of the vibrational modes in amorphous silicon.
Even for crystalline materials, it is well known that the lattice
modes are never completely harmonic, and anharmonic terms always
appear to higher order when expanding in the amplitude of the lattice
vibrations \cite{sr90}.
 Thus, the contribution of the anharmonic terms becomes
relevant as the phonon coordinate becomes large at finite temperatures.
A quantitative estimation of the overall anharmonicity
of the atom vibrations can be obtained from our PI MC
simulations by the
ratio between kinetic and potential energy at different temperatures.
According to the virial theorem, this ratio should be 1 for a
harmonic solid. In fig.\,5 we present the temperature
 dependence of this energy ratio for crystalline and amorphous
silicon (open and black symbols, respectively),
as derived from the quantum simulations.
As expected, the anharmonicity is found to
be higher for a-Si, although the kinetic-to-potential energy
ratio goes to 1 (within our statistical noise) at low $T$.
For the crystalline material, this energy ratio is close to 1
up to about 200 K, and at higher $T$ it decreases with a slope
smaller than in the case of a-Si.  Note that at the scale of fig.\,5,
the anharmonicity of the low-energy vibrational modes typical
of amorphous materials \cite{bu88,gi93} is not observable at low $T$.
Thus, the enhanced quantum delocalization in a-Si with respect to
c-Si is due (i) to a softening of the vibrational modes in
the material, as found from the vibrational density
of states for a-Si, even in a harmonic approximation; 
and (ii) to a larger anharmonicity of the vibrations
in the amorphous solid, as expected from the metastability
of a-Si. The resulting  quantum delocalization
is not negligible versus the structural disorder (described
in classical terms), as seen for the RDF in figs.\,1 and 2.

\begin{figure}
\centerline{\psfig{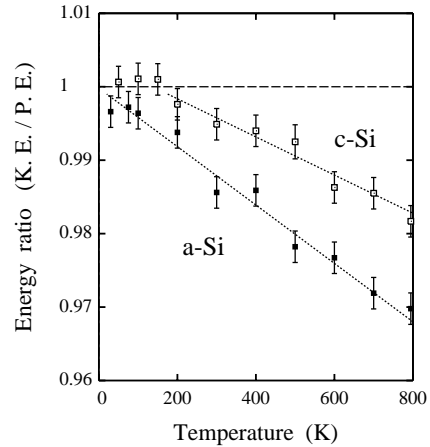}}
\vspace*{-2.5cm}
\caption{
Kinetic-to-potential energy ratio for crystalline (open squares) and
amorphous silicon (black squares), as derived from PI MC simulations.
Lines are guides to the eye. In a harmonic approximation, this energy
ratio equals 1 for all temperatures (dashed line).
}
\label{f5}
\end{figure}

In summary, path-integral Monte Carlo simulations provide us with 
a good tool to study quantum effects in disordered materials. 
Quantum effects are significant for structural observables such as the
radial distribution function of amorphous silicon, especially
at low temperatures.
Although quantitative values found by using the Stillinger-Weber
potential for the interatomic interaction can change by
employing other potentials, the main conclusions presented here
can hardly depend on the potential employed in the simulations.
 Quantum simulations similar to those presented here can
give information on the atom delocalization and anharmonic effects
in other amorphous materials. In particular, it is expected
that this method will give valuable information on
hydrogenated a-Si, where the quantum delocalization of hydrogen
can be crucial to understand the macroscopic properties of this
material.
In fact, nontrivial quantum effects have already been presented from
PI MC simulations of hydrogen in c-Si \cite{ra94}.
Also, quantum simulations in the context of the quantum
transition-state theory can give valuable information on the
low-energy tunneling states that control the thermal,
 acoustic, and dielectric properties of amorphous materials
at low temperatures.

This work was supported by CICYT (Spain) under Contract No.\,PB96-0874.
Helpful discussions with R. Ram\'{\i}rez and J. C. Noya are 
acknowledged.  M. A. Ramos and J. A. Verg\'es are thanked 
for critically reading the manuscript.

\end{document}